# The effects of graphical interface design characteristics on human-computer interaction task efficiency

R. MICHALSKI†, J. GROBELNY† and W. KARWOWSKI‡

†Institute of Industrial Engineering and Management (I23),
Faculty of Computer Science and Management (W8)
Wrocław University of Technology,
27 Wybrzeże Wyspiańskiego,
50-370 Wrocław, POLAND
e-mail: Rafal.Michalski@pwr.wroc.pl
e-mail: Jerzy.Grobelny@pwr.wroc.pl

‡Center for Industrial Ergonomics
University of Louisville
Lutz Hall, Room 445, Warnock Street
Louisville, Kentucky 40292, USA
e-mail: Karwowski@louisville.edu

Corresponding author:
Dr. Waldemar Karwowski
Center for Industrial Ergonomics
University of Louisville
Lutz Hall, Room 445, Warnock Street
Louisville, Kentucky 40292, USA
e-mail: Karwowski@louisville.edu
phone: +1 502 852 7173
fax:     +1 502 852 7397

# 1. Abstract

The main objective of this paper was to investigate the effects of a computer screen interface design and its related geometrical characteristics of 36 graphical objects on a user's task efficiency. A total of 490 subjects took part in laboratory experiments that focused on the direct manipulation of a visual dialogue between a user and a computer. The subjects were asked to select an object from among a group of items randomly placed on the computer screen that were visible exclusively during the visual search process. A model expressing the mean object acquisition time as a function of graphical object size and the configuration was developed and statistically validated. The model showed an influence of geometrical design characteristics of the graphical objects (icons) and their groupings (icon structures) on the observed task efficiency. The reported results can be used at those stages of a software lifecycle that concentrate on prototyping, designing, and implementing graphical solutions for the efficient graphical user-computer interface.

# 2. Introduction

Computer users have contact with an information system only with the help of an interface that defines information flow rules between a human and a machine.. The type of human-computer dialogue within an one-user interface is determined by the application of interactions. Usually, basic interaction styles include (Preece et al. 2002, Torres 2002, and Dix et al. 2004): questions and answers, command line languages, filling forms, menus, direct manipulation, and natural language. Innovative solutions have been suggested (Hartson 1998, Whittaker et al. 2000, Preece et al. 2002, Christou and Jacob 2003), such as ubiquitous computing, pervasive computing, wearable computing or wearables, virtual reality or environment, augmented reality, and affective computing. Despite enormous technological progress that enables creating more and more advanced systems, it occurs that the usefulness of new interaction styles is to a large extent limited (Hartson 1998). Whittaker and colleagues (2000) also noticed that most scientific publications in the human-computer interaction (HCI) field relate to modern interaction styles, whereas many research areas and opportunities of improving models, methods, and tools in traditional ways of human-computer communication are often ignored. Meanwhile, in contemporary IT systems, users most frequently exchange information with the application by pointing various graphical elements and by confirming (usually by clicking) the execution of a given activity. The 'point and click' method is a part of the human-computer interaction style called direct manipulation (Shneiderman 1982, 1983). In this kind of interaction, it is necessary for the user to use one of many available devices, for instance, light pens, digitizers, joysticks, arrow keys, track balls, touch screens and most of all, computer mice (Greenstein and Arnaut 1988).

# 3. Related research

The research concerned with the graphical interface design characteristics and their impact on human-computer interaction task efficiency can generally be divided into three main trends. The first trend involves investigations of the movement time in the 'point and click' method, where the user task is to select by means of any pointing device a given graphical object while it is constantly visible. In this case the problem of finding the item does not exist and the measured time expresses only the visually controlled motor activity.

The second trend includes studies in which the user has to search for a particular graphical item among the group of distractors in various conditions without the necessity of directly pointing the specified object. And finally, the third trend involves a combination of the first two groups. This time in order to complete a task one needs to identify and select the specified item by pointing and clicking on it. In all of the trends the efficiency can be measured by the time needed to complete the task and a number of errors accumulated. When the second and third trends are concerned with distractor influence, data gathered by the eye tracking systems may be applied e.g. number of fixations or fixation duration. However, it should be noted that in the work of Murata and Furukawa (2005) it was shown that search time cannot be solely explained by the variables related to eye movements, and that the display characteristics significantly influence reaction time either.

### 3.1. Movement time in the 'point and click' method (selecting graphical objects)

The basic research of features and tools regarding object selection carried out by the 'point and click' method started in the 1960's. It was shown that the most efficient tool in this technique regarding operation time and number of errors made is a computer mouse (English et al. 1967). In the 1970's, the classical Fitts' law[*] was used to describe the process of pointing graphical objects on the computer screen (Fitts 1954, Fitts and Peterson 1964). The suitability of this approach in relation to elements constantly visible during visual search process was confirmed in numerous subsequent studies (Card et al. 1978, Epps 1986). A comprehensive review of studies concerning Fitts' law application for various pointing devices, along with an in-depth discussion presented by MacKenzie (1991, 1992) and extensive analysis of its use in many fields, can be found in the work of Plamondon and Alimi (1997). This HCI research trend has lasted until now and is concerned with the utilization of Fitts' law, its modifications, and the improvements in standard and new interaction styles (MacKenzie 1995a and 1995b, Coninx et al. 1997, Murata and Iwase 2001, Sallnäs and Zhai 2003, Blanch et al. 2004, Guiard and Beaudouin-Lafon 2004, Balakrishnan 2004, Zhai 2004, Zhai et al. 2004). The most recent works regarding the application of Fitts' formula in HCI are reported in the article of Soukoreff and MacKenzie (2004).

### 3.2. Visual search in graphical interface design

The problem of choosing graphical objects on the computer screen cannot always be described solely by means of Fitts' law. Apart from the visually controlled motor activities, there are also cognitive components that may have an impact on the acquisition times in the human-computer interaction (Card et al. 1983). Although there were numerous studies on visual search aimed at properly identifying the target item, one of the first investigations specifically focused on determining the impact of computer interface features on the visual search task efficiency was carried out by Backs and colleagues (1987). They asked eight operators to perform visual search tasks. The subjects had to find a target object in vertical and horizontal menus containing 4, 8 or 12 items and report an associated numerical value. The authors found that search time was shorter for the vertical than for horizontal configurations. The menu size also significantly influenced the response time. The smaller number of items was there in the menu the shorter search time was observed. The visual search for the target letter X in five different configurations of randomly generated alphabetic (upper case) arrays was studied by Scott and Findlay (1991). The results obtained from the twenty participants showed that the response time was significantly higher in the vertical array comprising of 100 rows and 25 columns than in the horizontal configuration 25 characters high and 100 characters wide. Other studies carried out on the group of 14 volunteers by Niemelä and Saarinen (2000) regarded the effect of icon presence and their grouping on the process of searching a file in a computer graphical interface. The obtained results indicate that both examined factors have positive and statistically significant influence on the speed of accomplishing the task. Similar experiments with slightly different icons were also reported in the works of Murata et al. (2002) and Murata & Furukawa (2005). They used an eye tracking device for gathering data from ten volunteers. The analysis provided similar results to those demonstrated in the original study.

---

[*]Classical Fitts' law says that visually controlled target movement time ($MT$) depends on the target object width ($W$) and the movement amplitude ($A$). $MT = a + b \times ID$, where $a$ and $b$ parameters are obtained experimentally, while $ID = \log_2(2 \times A / W)$.

A number of studies with equipment for tracking and registering eyeball movements was applied to assess how display geometrical features and some other conditions affect information retrieval time were also carried out. For instance, Näsänen and colleagues (2001a) focused on investigating the influence of image quality (low and high), screen type (CRT and LCD) and font sizes (Courier New 8, 11, 16, 23, 35 pt) on the acquisition times. For small character sizes at low picture quality, the speed of visual search was lower for LCD monitors. For high quality images, the differences in finding efficiency for both monitor types were insignificant. Similar studies were carried out by Näsänen and colleagues (2001b). They analyzed visual search times of Latin characters arranged in a square matrix, depending on the element's contrast and matrix size ($3 \times 3$, $5 \times 5$, $7 \times 7$, and $10 \times 10$). The authors showed that along with increasing the contrast level, the efficiency of finding the appropriate object grew for all examined sets of icons. Lindberg and Näsänen (2003) carried out experiments aimed at identifying the influence of computer icon set sizes (squares $2 \times 2$, $3 \times 3$, $7 \times 7$, $10 \times 10$) and various spacing between elements on panels (0, 8, 16, 32, 64 pixels) on times needed for finding a given object within examined graphical structure. They obtained results that confirmed that the matrix size has a meaningful impact on search times, but the distance between icons was of no importance. In turn, Näsänen and Ojanpää (2003) utilized similar icons in square configurations ($13 \times 13$) to analyze the effects of graphical object's contrast and sharpness on the speed of searching for a specified icon. They reported that the increase in contrast, or sharpness, improved visual search efficiency. This effect was observable only up to the medium level of the parameters, further rise in object quality did not affect the search time.

There were also some works related to spatial menu layout in web pages and its impact on the visual search performance. For example, Schaik and Ling (2001) examined the effect of navigation menu layout (top, bottom, left and right) containing five hypertext links and background contrast on visual search performance in mock web pages. The experiments were completed by one hundred and eighty nine undergraduates. The authors demonstrated that the speed of executing the task was significantly higher for the menus located at the top or left of the screen while reaction time was not affected by the contrast. However Pearson and Schaik (2003) in a very similar experiment carried out on forty participants obtained somewhat different results. They investigated the effect of navigation menu layout and hypertext link color (blue and red) on information retrieval efficiency. The authors used the same number of layout items and types of menu locations as in the work of Schaik and Ling (2001). The link position factor significantly influenced reactions times while the effect of link color was supported. There was no difference in reaction times between left and right conditions and between top and bottom positions. However, generally horizontal locations were searched faster than vertical ones.

### 3.3. Studies in HCI related to movement time and visual search

The relationship between visually controlled motor activities described by the Fitts' low and cognitive factors involved in the visual search process does not necessarily exhibit an additive nature. For example, during the search task one may at the same time move the pointer towards the right direction shortening the overall task completion time. Because the process of identifying and selecting objects in graphical computer interfaces is complex, it is justified to carry out research encompassing both of the conditions simultaneously.

A number of studies on combined manual-decision tasks can be found in the literature, but only few of them were focused on the interaction between a human and a computer. A comprehensive review and re-analysis of the research in other than HCI areas was presented by Hoffmann and Lim (1997). They found that for the sequential tasks the additive form of the choice and movement time was most appropriate but in the case of concurrent tasks, the motor and cognitive activities interfered. One of the first studies concerned with various design configurations and efficiency measures including simultaneously visual search and movement time were studies described in the paper of Deininger (1960). In a series of experiments he investigated the performance of keying telephone numbers using sixteen various arrangements of ten numerical keys. There were no statistically significant differences with respect to the keying time and error rates among standard rotary dial and four best configurations: horizontal (2 rows × 5 columns), vertical (5 × 2), square (3 × 3 plus 1), and the rotary arrangement similar to the standard one but turned round in the right direction by 270 degrees. The conclusion was that any of these five arrangements were acceptable in terms of the operation efficiency, but the vertical layout should be avoided because it was disliked by many subjects. Many subsequent studies regarding keying issues in physical and virtual keypads or keyboards carried out before 1999 were comprehensively reviewed in the work of Kroemer (2001).

The research on selecting an element from among a group of virtual objects and the influence of geometrical factors of such sets on the acquisition times were probably initiated by Drury and Hoffmann (1992). These authors showed that for simulation on computer screen keyboards with given spaces between keys, the optimal key width occurs when the space between adjacent buttons equals the finger pad[*] width. Succeeding studies were connected with different sizes and configurations of virtual keyboards presented on various types of screens (e.g. Sears et al. 1993, MacKenzie and Zhang 1999, MacKenzie et al. 1999, Sears et al. 2001, MacKenzie and Zhang 2001, Zhai et al. 2002, Sears and Zha 2003, Lee and Zhai 2004). In most cases, the keys' arrangements and/or their sizes were significant factors influencing the operation efficiency.

In addition to the studies related to keypads and keyboards there were also some research focused directly on graphical computer interface features. For example, Shih and Goonetilleke (1998) investigated the efficiency of identifying and selecting computer buttons from vertical and horizontal menus consisting of ten items in Chinese and English languages. The experiments were conducted on fifty Hong Kong Chinese students. The obtained results indicated that the horizontal menu was operated more quickly than the vertical one, regardless of whether English or Chinese texts were used on the buttons. In other studies Goldberg and Kotval (1999) compared two sets of eleven computer buttons used commonly in graphical programs. The first collection was functionally grouped, while in the second one icons were placed randomly. Researchers employed eyeball movement measures to analyse data gathered during experiments made on twelve subjects and proved that the functionally grouped set of icons was operated more efficiently.

---

[*]The part of a fingertip which directly touches a keyboard button.

In turn, Fleetwood and Byrne (2002) conducted an experiment on a group of 20 persons, analyzing the influence of an icon number (6, 12, 18, 24), border type (square, circle, and without a border), and quality (poor, medium, and good) on the time needed for selecting the specified icon. They found that the effect of icon quantity and quality considerably influenced search mean times, whereas the effect of the icon border type was statistically insignificant. The same authors replicated their experiments (2003) on ten subjects broadening the analysis by examination of data obtained from the eye movement registering system. Apart from confirming the previous results, they also observed the color of good quality icons was helpful during the visual search process. Everett and Byrne (2004) used icons identical to those examined in articles of Fleetwood and Byrne (2002 and 2003) to study the influence of set sizes, icon quality, and spacing on acquisition times. The results showed a significant impact of all three elements on the dependent variable. The detailed description and comparative analysis of experiments carried out by Fleetwood and Byrne (2002 and 2003), as well as Everett and Byrne (2004), are reported in the work of Fleetwood and Byrne (in press). An interesting study regarding the impact of spatial layout of graphical objects on the efficiency of visual search and selecting the item by means of a computer mouse was carried out by Simonin and colleagues (2005). They analyzed thirty realistic color photos arranged along four different structures: pictures randomly placed on a computer screen, array consisting of five rows and six columns, two concentric ellipses and, eight radii along medians and diagonals of the screen. The task performances for all five participants happened to be independent of the display layout.

The article published by Grobelny and colleagues (2005) is one of the latest works in this area. The authors examined the influence of geometrical parameters on the efficiency of the 'point and click' method in the MS Windows™ graphical environment. The obtained results showed that in the case of icons constantly visible on the screen, the mean acquisition time for one element can be described by means of Fitts' classical law; whereas for structures showed only during the selection time, the configuration of icon sets had to be taken into consideration. The superiority of square arrangements over horizontal and vertical arrangements was also reported. The results suggested that for sets consisting of 16 elements or more there was an optimal size of graphical objects. Grobelny and colleagues also proposed formulas describing the relationship between mean acquisition times for analyzed panels containing 8 and 16 icons, and for geometrical factors of the structures and their components.

## 4. Research objectives

In the present study we are continuing research related to both movement time and visual search tasks in human-computer communication. Previous studies provided some insights with respect to the impact of various design characteristics on users' performance. However, in some aspects the obtained results are not consistent especially when the display configuration is concerned. Therefore, the first objective of this research was to provide additional insight into the problem of designing efficient graphical panels. Although, search tasks and experimental procedures used in our experiments were similar to those employed in experiments reported by Grobelny and colleagues (2005), we substantially broadened the range of various panel design characteristics. In the same paper it was also shown that the Fitts' formula is inadequate in this type of search tasks and a different model was proposed. Because we analyzed a greater number of panel configurations that contained a considerably larger number of items, the additional purpose of this paper was to provide a formal model explaining average operation time by means of the graphical features of tested structures. We also explain our findings and provide guidelines for designing panels of graphical items in HCI based on the obtained empirical evidence.

## 5. Method

### 5.1. Participants

A total of 490 volunteers participated in the study. All reported having a normal or corrected to normal visual acuity. Most of the subjects attended the first, second, third, or fourth year of the full-time Master's program (270 people, 55%), and the remaining 220 were from the first and second years of a part-time complementary Master's study organized by the Computer Science and Management Faculty at the Wroclaw University of Technology. There were fewer female participants (234 subjects, 48%) than males (256 subjects, 52%). The majority of subjects (348) were young students within the age range of 21-25 years old; students who did not exceed 30 years of age accounted for about 93% of the total number of participants. Almost all (97%) of the volunteers worked with computers at least several times a week, and a substantial majority (78%) used computers on a daily basis. Only two people worked with computers once a month or rarely. Users most frequently spent from two to three hours a day at a computer. The majority of participants (76%) used computers, on average, no longer than 6 hours a day; however, there were 24 people (5%) who spent more than 10 hours a day working at a monitor screen. The number of hours per day working with computers included only the time spent on actively using the software. Activities such as watching TV on the computer screen or listening to the music played by appropriate programs were not taken into consideration.

### 5.2. Apparatus

A special-purpose computer application was designed and developed to carry out the experiments. The software was written in a MS Visual Basic™ 6.0 environment and was based on a relational MS Access™ database. All necessary information regarding the studies (e.g. questionnaire answers and task parameters, such as acquisition times or errors made) was collected in a database file. A function that enabled all importing data from different workstations into one file was also implemented in the program. Experimental results were gathered in one place allowing for direct record transfer to information systems supporting statistical analysis (e.g. SPSS™, Statistica™, and Statgraphics™) by means of ODBC[*] technology. The research was carried out in teaching laboratories on uniform personal computers equipped with identical optical computer mice and 17" monitors of the CRT type. On all computers screens, the resolution was set at 1024 by 768 pixels and a typical (default) computer mouse parameters were applied.

### 5.3. Independent variables

Three independent variables differentiating the analyzed graphical object structures were manipulated:

    *Graphical object size*. The standard square buttons used in Microsoft[®] operating systems were employed. The side button sizes in TWIPs[†], pixels, millimetres and approximate

---

*Open DataBase Connectivity.
[†]At 1024 by 768 screen resolution, 1 pixel amounts to 15 TWIPs.

visual angle are presented in Table 1. These objects sizes are commonly utilized in many computer programs, e.g. in a popular MS Office™ package or in such applications as AutoCAD™, Corel Draw™, and 3d Studio Max™. The choice of a square shape of elements on one hand simplified the experiments, but was also rational in light of Martin's (1988) research; he found that the square keys presented on touch screens were the best from among examined rectangular button shapes.

*Panel location on the screen.* Two locations were investigated: the upper left and the upper right screen corners. The arrangements were moved away from the screen edges by 270 TWIPs in order to minimize the effect of easier selection of icons placed on the external borders of the sets. The distance was equalled to the height of the top title bar present in most dialogue windows used in Microsoft® operating systems.

*Panel configuration.* Seven types of arrangements were analysed: squares with a side size of six elements (06_06); three types of horizontal rectangles: two rows and 18 columns (02_18), three rows and 12 columns (03_12), and four rows and nine columns (04_09); and three types of vertical rectangles: 18 rows and two columns (18_02), 12 rows and three columns (12_03), and nine rows and four columns (09_04). Exemplary arrangements of studied graphical structures are illustrated in Figure 1.

Table 1. Sizes of graphical elements used in experiments.

| Name | TWIPs | Pixels | Millimetres | Visual angles |
|---|---|---|---|---|
| Small | 330 | 22 | 6 | 0°41' |
| Medium | 450 | 30 | 8 | 0°55' |
| Large | 570 | 38 | 10 | 0°69' |

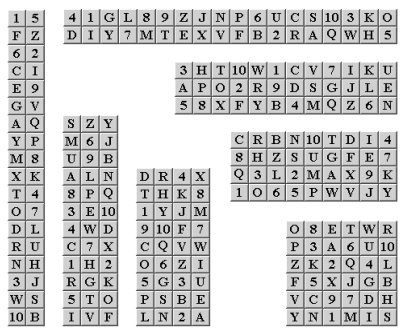

Figure 1. Examples of studied arrangements.

## 5.4. Dependent measures

The dependent variables being measured were the acquisition time and the number of errors made. The time was computed from when the START button was pressed, to when the icon was selected. The error occurred when the user selected different than required graphical object, however the user did not received immediate feedback about the mistake.

## 5.5. Experimental design

Three independent variables resulted in forty two different experimental conditions: (three object sizes) × (two panel locations) × (seven configurations). A non-standard mixed model design (between and within subjects) was used to investigate all of the 42 sets of objects.

The employed design is connected with our forthcoming studies and requires some explanation. The ongoing research includes assessment of users' preferences regarding the same graphical panels and the relationship between objective efficiency measures and subjective ratings. We employ the Analytic Hierarchy Process (AHP), which requires pairwise comparisons and provides only relative vectors of preferences (Saaty 1980), and some form of conjoint analysis (Luce and Tukey 1964) for examining relationships between efficiency and subjective measures. Therefore, in order to be able later to analyze preferences for different panel features, we decided to assign the conditions to six experimental groups in a way demonstrated in Table 2.

Table 2. Experimental conditions.

| | Left | | | | | | | | | | | | | | | | | | | | | Right | | | | | | | | | | | | | | | | | | | | |
|---|---|---|---|---|---|---|---|---|---|---|---|---|---|---|---|---|---|---|---|---|---|---|---|---|---|---|---|---|---|---|---|---|---|---|---|---|---|---|---|---|---|---|
| | Small | | | | | | | Medium | | | | | | | Large | | | | | | | Small | | | | | | | Medium | | | | | | | Large | | | | | | |
| | 02_18 | 03_12 | 04_09 | 06_06 | 09_04 | 12_03 | 18_02 | 02_18 | 03_12 | 04_09 | 06_06 | 09_04 | 12_03 | 18_02 | 02_18 | 03_12 | 04_09 | 06_06 | 09_04 | 12_03 | 18_02 | 02_18 | 03_12 | 04_09 | 06_06 | 09_04 | 12_03 | 18_02 | 02_18 | 03_12 | 04_09 | 06_06 | 09_04 | 12_03 | 18_02 | 02_18 | 03_12 | 04_09 | 06_06 | 09_04 | 12_03 | 18_02 |
| G1 | ○ | × | × | × | × | × | ○ | ○ | × | × | × | × | ○ | × | ○ | × | × | × | ○ | × | ○ | ○ | × | × | × | ○ | × | ○ | ○ | ○ | ○ | ○ | ○ | ○ | ○ | ○ | ○ | ○ | ○ | ○ | ○ | ○ |
| G2 | × | ○ | ○ | ○ | ○ | ○ | ○ | ○ | ○ | ○ | ○ | ○ | ○ | ○ | ○ | ○ | ○ | ○ | ○ | ○ | ○ | ○ | × | × | × | × | ○ | × | × | ○ | × | × | × | ○ | × | × | ○ | × | × | ○ | × | × |
| G3 | ○ | ○ | ○ | ○ | ○ | ○ | ○ | × | ○ | × | × | ○ | × | × | ○ | ○ | ○ | ○ | ○ | ○ | ○ | ○ | × | × | × | × | ○ | × | × | ○ | × | × | × | ○ | × | ○ | ○ | ○ | ○ | ○ | ○ | ○ |
| G4 | ○ | ○ | ○ | ○ | ○ | ○ | ○ | ○ | ○ | ○ | ○ | ○ | ○ | ○ | × | × | ○ | ○ | × | × | × | ○ | ○ | ○ | ○ | ○ | ○ | ○ | ○ | ○ | ○ | ○ | ○ | ○ | ○ | × | × | ○ | ○ | × | × | × |
| G5 | × | × | ○ | ○ | × | ○ | × | × | × | ○ | × | ○ | × | ○ | ○ | ○ | ○ | ○ | ○ | ○ | ○ | ○ | ○ | ○ | ○ | ○ | ○ | ○ | ○ | ○ | ○ | ○ | ○ | ○ | ○ | ○ | ○ | ○ | ○ | ○ | ○ | ○ |
| G6 | ○ | ○ | ○ | ○ | ○ | ○ | ○ | ○ | ○ | ○ | ○ | ○ | ○ | ○ | ○ | ○ | ○ | ○ | ○ | ○ | ○ | × | ○ | × | × | × | ○ | × | ○ | ○ | ○ | ○ | ○ | ○ | ○ | × | ○ | × | × | ○ | × | × |

× = condition used; ○ = condition not used

The subjects were randomly assigned to groups: in the first group (G1) there were 84 participants; 87 in G2; 81 in G3, G4, G5 each; and 76 persons in G6. The demographical and computer literacy characteristics were not controlled – the questionnaires were filled out after the assignment to the groups took place. Because of that, it was checked whether there were any differences in assignment to groups based on these variables. The employed *Chi-square* tests did not reject any of hypotheses that the occurrences of a given characteristic in individual groups were equal to occurrences in the whole sample (Table 3).

Table 3. Differences in assignment to groups, based on demographic variables.

| Factor | $df$ | $\chi^2$ | $p$ |
|---|---|---|---|
| Gender (men, women) | 5 | 8.70 | 0.12 |
| Study type (full-time, part-time) | 5 | 2.64 | 0.76 |
| Age in years (16–25, 26–30, >30) | 10 | 10.0 | 0.76 |
| Computer possession in years (<3, 3–5, >6) | 10 | 14.4 | 0.16 |
| Frequency of computer usage (Everyday, Several times a week, Less than several times a week) | 10 | 14.7 | 0.14 |

All of the analyzed panels consisted of 36 identical buttons. Icons representing 26 Latin alphabet characters and ten Arabic numbers were placed on these buttons. Bolded Times New Roman font types in three different sizes: 12, 18 and 24 pt were used. In order to avoid potential mistakes between the O character and a 0 (zero) digit, numbers from 1 to 10 were utilized. The subjects were asked to select a graphical object from among a group of randomly placed items, which were visible exclusively during the visual search process. The effects of learning were not examined, and all 'search and select' tasks were executed by means of a standard computer mouse. The distance between the user and the computer monitor was set approximately at 50 cm. The visual angles of all the examined layouts are listed in Table 4.

Table 4. Panel sizes in visual angles (degrees).

| Panel arrangement | Small icons | | Medium icons | | Large icons | |
|---|---|---|---|---|---|---|
| | Width | Height | Width | Height | Width | Height |
| 02_18 | 12.9 | 1.4 | 17.3 | 1.9 | 21.7 | 2.4 |
| 03_12 | 8.6 | 2.2 | 11.6 | 2.9 | 14.5 | 3.7 |
| 04_09 | 6.5 | 2.9 | 8.7 | 3.9 | 10.9 | 4.9 |
| 06_06 | 4.3 | 4.3 | 5.8 | 5.8 | 7.3 | 7.3 |
| 09_04 | 2.9 | 6.5 | 3.9 | 8.7 | 4.9 | 10.9 |
| 12_03 | 2.2 | 8.6 | 2.9 | 11.6 | 3.7 | 14.5 |
| 18_02 | 1.4 | 12.9 | 1.9 | 17.3 | 2.4 | 21.7 |

### 5.6. Procedure

Participants were informed about a goal and a detailed range of the research. Subjects were also informed about the opportunity of receiving their results in the form of text file generated by the software in order to increase their motivation to participate. The file could be viewed on the spot or sent by e-mail to the volunteer following the experiment. The study began by filling out a questionnaire regarding personal data and computer literacy. Just before the experiment, each subject executed a test task. In this mode the results were not registered. Attemptive trials were performed until the volunteer stated that he was ready to do the real tasks. Then a dialogue window appeared with a START button and an order of searching a specified object (Figure 2) – and at this moment the tested graphical structure was not visible. When a user clicked the START button, the instruction window disappeared and one of the examined panels of items was shown.

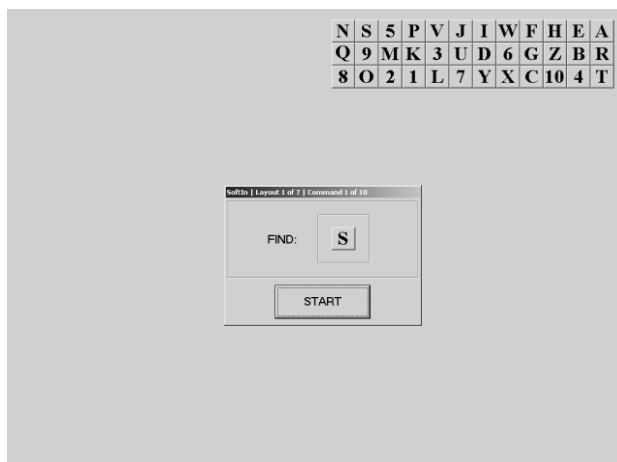

Figure 2. Exemplary set of icons and an order dialogue window.

Subjects had to find and select the earlier presented graphical item as fast as possible. The START button appeared for each trial so every time, the participant had to click START first and then the searched item. During executing orders, only research software was visible and other applications or operating system control elements were not available. Accidental selection of any button, other than the START button, resulted in presenting appropriate guiding information. Volunteers were allowed to execute the task exclusively by means of a computer mouse, and all keyboard buttons and shortcut keys were programically disabled. For every set of icons there were ten execution orders with a randomly chosen button. Once shown, an element could appear again. The order of presenting an individual group of items was randomly set for every participant. The location of every graphical object within the given set was also specified at random. The task of icon selection was repeated 10 times for each of the analyzed structures. Information about obtained mean acquisition times and incorrect attempts was shown every 10 trials. The average times were calculated only for properly selected elements.

## 6. Results

### 6.1. Descriptive statistics

Basic descriptive statistical data regarding obtained acquisition times are presented in Tables 5, 6, and 7. The parameters were grouped in three main categories: central tendency measures, variability measures, and shape characteristics. In many commonly used statistical methods, such as the analysis of variance or regression analysis, it is usually assumed that the examined dependent variables have normal distribution. As a result, it would be favorable if a variate defining acquisition times had a Gaussian distribution. Analyzing the data in Table 5, it can be noted that the median, mode, and mean values differ one from another (mode = 1422 ms, median = 1682 ms, and mean = 2189 ms). The mode is as much as 54% smaller than the mean value. In the case of the Gaussian distribution, these three parameters should be comparable. The calculated skewness amounted to 4.64, which is considerably different from the zero skewness characteristic of symmetric distributions, including of course the normal distribution. The positive sign of the skewness denotes that most of the variate values are located on the left hand side of the distribution. The kurtosis value (50.8) calculated from the sample is significantly different from the zero value of the typical kurtosis for the Gaussian distribution. The far bigger kurtosis value estimated from the empirical data suggests that the probability density distribution of the analyzed variate is considerably less dispersed than the normal distribution.

Table 5. Central tendency measures of acquisition times.

| Central tendency measures | |
|---|---|
| Mean | 2 189 ms |
| Geometric mean | 1 814 ms |
| Harmonic mean | 1 576 ms |
| Median | 1 682 ms |
| Mode | 1 422 ms |

Table 6. Variability measures of acquisition times.

| Variability measures | |
| --- | --- |
| Minimum | 297 ms |
| Maximum | 52 896 ms |
| Standard deviation | 1 758 ms |
| Variance | 3 090 920 $(ms)^2$ |
| Variability coefficient | 80% |

Table 7. Shape characteristics of acquisition times.

| Shape characteristics | |
| --- | --- |
| Skewness | 4.64 |
| Kurtosis | 50.80 |

The analysis presented above was also supported by the graphical comparison of experimental density function and normal distribution with parameters calculated from the sample data. The Figure 3 clearly shows that empirical acquisition times do not correspond with the theoretical curve.

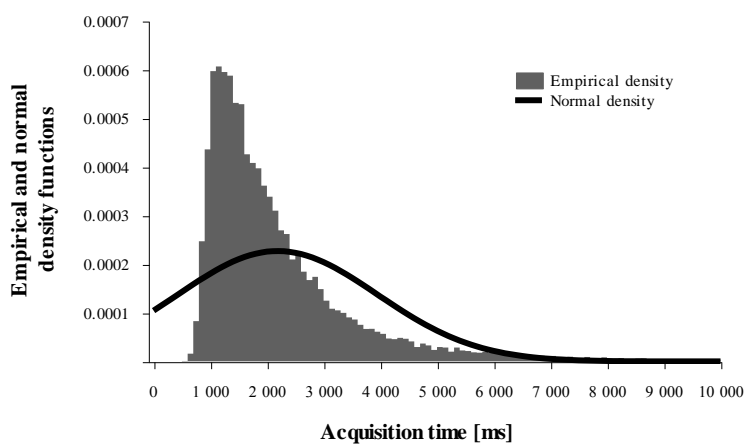

Figure 3. Experimental and normal density functions.

## 6.2. Analysis of variance

The analysis of variance was carried out by means of the Generalized Linear Models (GZLM) under the assumption that the dependent variable has the inverse Gaussian (IG) distribution. The GZLM was first defined by Nelder and Wedderburn (1972), and as opposed to General Linear Models (GLM), do not require a dependent variable to have a normal distribution. Furthermore, the assumption concerning constant variance of a random component does not have to be met. The GZLM includes a regression analysis as well as an analysis of variance. In both cases, statistical distribution of a dependent variable has to belong to a natural exponential family of distributions (e.g. normal, Poisson, binomial, gamma, or inverse Gaussian). In the work of Michalski (2005), it was shown that the hypothesis about the IG character of the acquisition time empirical distribution for the examined graphical structures cannot be rejected.

### 6.2.1. Geometrical factors

A three factorial analysis of variance based on the GZLM was also employed for assessing the effects of the panel location on the screen, the graphical object arrangement within the individual group, and item sizes. Means depending on the structure position did not differ considerably ($W^* = 0.9$, $df = 1$, $p = 0.34$), but average values for distinct arrangements were significantly different ($W = 129$, $df = 6$, $p < 0.0001$). Apart from that, the analysis indicated that there is a significant interaction between the location and set arrangement ($W = 18.1$, $df = 6$, $p < 0.006$). In further consideration, the relationship between the structure arrangement and the icon sizes was also taken into account because the assumed significance level ($\alpha = 0.05$) was only slightly exceeded ($p = 0.057$). Mean acquisition times for every analyzed arrangement are illustrated in Figure 4. From the presented data, it can be observed that configurations consisting of nine rows and four columns (2122 ms) together with six rows and six columns (2128 ms) were operated the fastest. The worst results were obtained for the arrangements comprising of two columns and eighteen rows (2344 ms) as well as three columns and twelve rows (2223 ms).

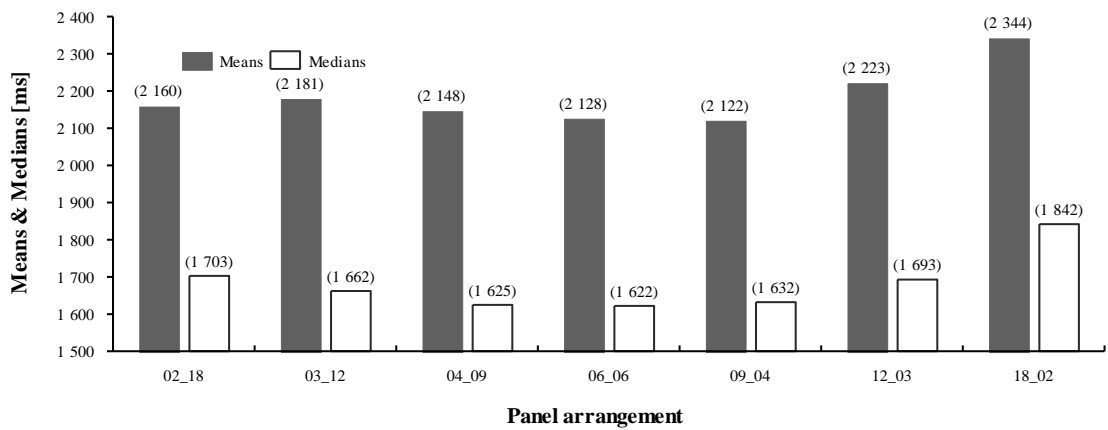

Figure 4. Means and medians of acquisition times depending on structure arrangements (W = 129, df = 6, p < 0.0001).

For the sake of the interaction between panel location and arrangement ($p < 0.006$), each type of configuration was examined to determine if the mean acquisition times differ depending on graphical structure position on the screen. The analysis results are shown in Table 8 and illustrated in Figure 5. The mean acquisition times for icon sets on the left were significantly different from the mean acquisition times obtained on the right for only two of the seven arrangements. The significance level for panels containing two columns and eighteen rows yielded a value of $p = 0.079$.

Table 8. Mean acquisition times depending on panel location and arrangement.

| Panel arrangement | Acquisition times (ms) | | Statistics ($df = 1$) |
|---|---|---|---|
| | Left | Right | |
| 09_04 | 2 196 | 2 069 | $W = 13.2$, $p < 0.0003$ |
| 12_03 | 2 178 | 2 283 | $W = 9.63$, $p < 0.0020$ |
| 18_02 | 2 379 | 2 317 | $W = 3.09$, $p = 0.0788$ |

[*]Wald statistics.

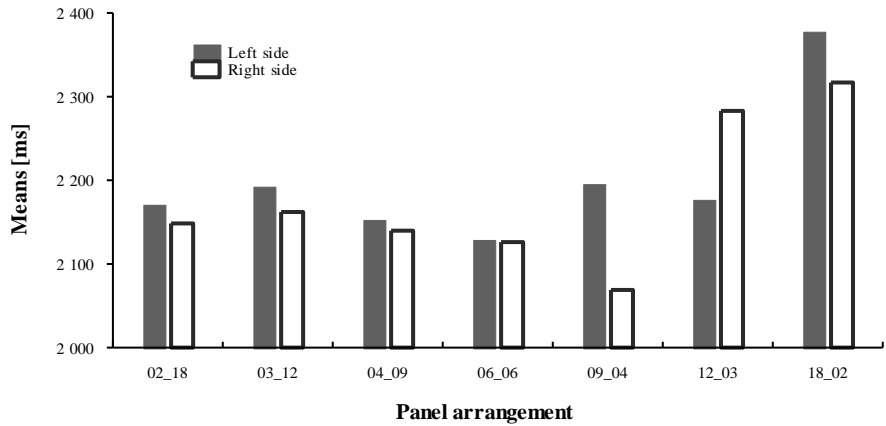

Figure 5. Mean acquisition times depending on the panel location and arrangement.

Meaningful discrepancies were observed exclusively in vertical groups of icons. In the case of 09_04 and 18_02 configurations, the sets located on the right were operated faster; however, arrangements comprised of three columns and twelve rows structures on the left hand side seemed to be better. All the square and horizontal panels were operated quicker on the righthand side, and the differences were statistically insignificant.

Results of the analysis regarding mean acquisition times for all examined arrangements in relation to icon sizes are presented in Table 9. The significance of the interaction was equal to $p = 0.057$. The one-way analysis of variance (GZLM) on three levels was carried out for each of the arrangements. The analysis of variance results show that the item size effect was significant for each panel arrangement. Graphical representations of values from Table 9 are placed in Figure 6. The tendency of decreasing mean acquisition times along with the rise of object dimensions can be easily observed. There is only one exception from this rule.The structures containing four rows and nine columns with panels of medium-sized items were operated faster than groups comprised of large graphical elements.

Table 9. Mean acquisition times depending on the panel arrangement and icon sizes.

| Panel arrangement | Acquisition times (ms) | | | Statistics ($df = 2$) |
|---|---|---|---|---|
| | Small | Medium | Large | |
| 02_18 | 2 307 | 2 165 | 2 058 | $W = 33.2$, $p < 0.0001$ |
| 03_12 | 2 314 | 2 165 | 2 091 | $W = 36.7$, $p < 0.0001$ |
| 04_09 | 2 244 | 2 078 | 2 158 | $W = 14.5$, $p < 0.0008$ |
| 06_06 | 2 283 | 2 091 | 2 021 | $W = 49.3$, $p < 0.0001$ |
| 09_04 | 2 298 | 2 125 | 2 001 | $W = 49.5$, $p < 0.0001$ |
| 12_03 | 2 397 | 2 193 | 2 161 | $W = 28.0$, $p < 0.0001$ |
| 18_02 | 2 505 | 2 315 | 2 272 | $W = 24.4$, $p < 0.0001$ |

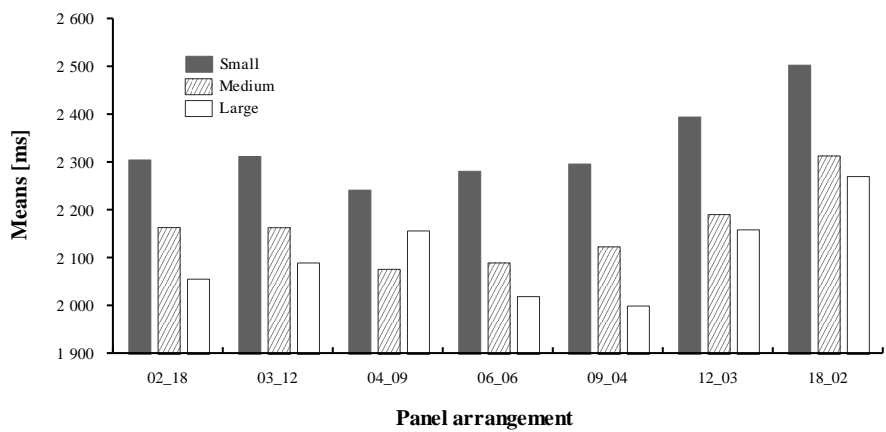

Figure 6. Mean acquisition times depending on the panel arrangement and icon sizes.

The three way analysis of variance based on GZLM was also employed to study the significance of differences between mean acquisition times depending on panel location on the screen (left or right hand side), orientation (horizontal, square, and vertical), and sizes of graphical objects incorporated in sets (small, medium, and large). Although in this case, the ANOVA was substantially unbalanced (as there was three times as many horizontal and vertical panels as square ones), facilitating rough comparisons with studies in which only one type of vertical or horizontal arrangements was employed (e.g. Backs et al. 1987, Grobelny et al. 2005). Apart from that, it gives some more insight in the obtained data. The obtained results were similar to the first ANOVA and proved that the structure location did not meaningfully influence the mean acquisition times ($W = 0.81$, $df = 1$, $p = 0.37$) while element size ($W = 175$, $df = 2$, $p < 0.0001$) had a significant effect. The orientation factor had a significant impact on the mean times ($W = 58.8$, $df = 2$, $p < 0.0001$) and none of the interactions between examined factors were statistically significant ($\alpha = 0.05$). Incorrect selections were not taken into account during the analysis.

Mean acquisition time values depending on the graphical structure orientation are presented in Figure 7. The average times were considerably shortest for sets containing objects grouped in squares. The relative difference between the mean acquisition times for squares and horizontal sets amounts to 1.7% and is statistically significant ($W = 5$, $df = 1$, $p = 0.031$). Users, on average, needed most time for task completion during testing of vertically oriented configurations. The biggest percentage discrepancy occurred between square and vertical sets and was equal to 5.1%.

Means and medians of acquisition times depending on set element sizes are shown in Figure 8. Structures containing large objects operated 2.5% faster than those with elements of medium size. The average acquisition times for sets that included medium icons were shorter by 7% from the results registered for configurations with small icons.

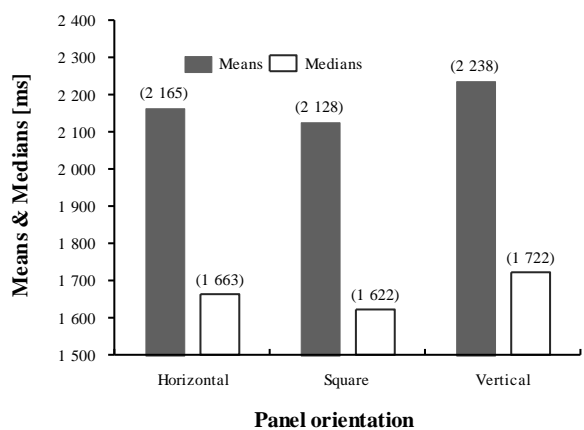

Figure 7. Means and medians of acquisition times depending on the structure orientation (W = 58.8, df = 2, p < 0.0001).

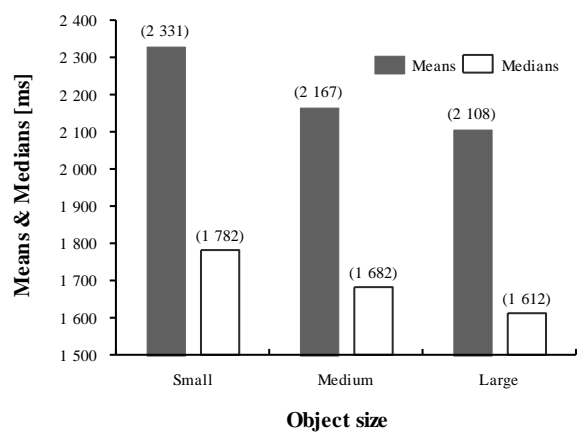

Figure 8. Means and medians of acquisition times depending on the icon sizes (W = 175, df = 2, p < 0.0001).

### 6.2.2. Demographical factors and computer literacy.

Taking advantage of the information gathered with the help of questionnaires, many analyses concerning changes in mean acquisition times obtained by participants from specified categories were conducted. Classifications connected with demographical data as well as factors linked with computer literacy were considered. Results of panel operation speeds conditional on the user gender demonstrated that female performance (2148 ms) was much better ($W = 39.3$, $df = 1$, $p < 0.0001$) than male performance (2226 ms). The percentage difference amounted to 3.5%. Means and medians of acquisition times subject to the participant's age are presented in Figure 9, where a trend of speed decrease for older subjects can be observed. ANOVAs results regarding comparisons of mean acquisition times depending on computer competencies included a computer possession period, frequency of PC usage, and daily amount of time devoted to working with a computer. Average times and medians pertaining to the computer possession period are illustrated in Figure 10. The longer the subjects owned the PC, the higher the operation speed of examined configurations. The discrepancy between average results for extreme categories was equal to 6%. The effect of computer usage frequency had a considerable impact on panel operation performance. In Figure 11, the tendency of decreasing mean acquisition times along with the increase of frequency of PC usage can be observed.

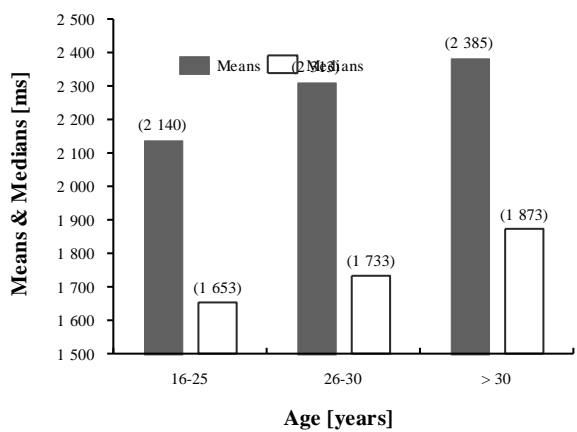

Figure 9. Means and medians of acquisition times depending on user's age (W = 174, df = 2, p < 0.0001).

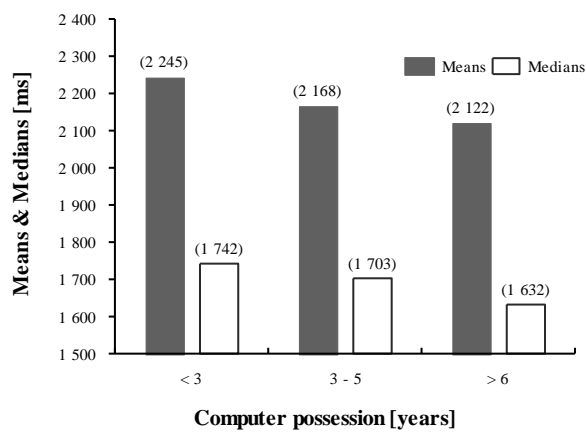

Figure 10. Means and medians of acquisition times depending on computer possession (W = 64.5, df = 2, p < 0.0001).

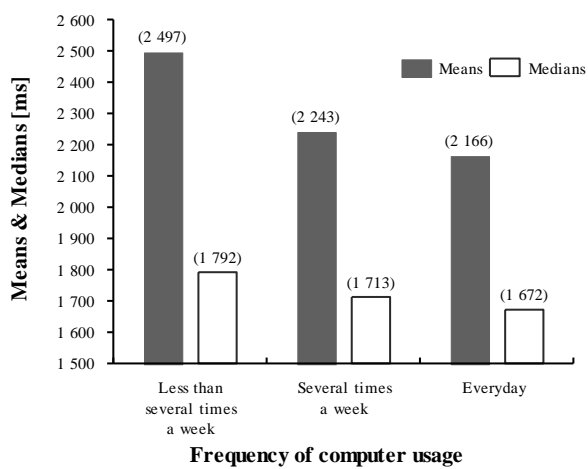

Figure 11. Means and medians of acquisition times depending on frequency of computer usage (W = 75.7, df = 2, p < 0.0001).

A statistically significant difference ($W = 19.2$, $df = 1$, $p < 0.0001$) was also observed between people who spent less than three hours daily in front of a computer (215 people, 2220 ms), and those who used a PC more than three hours a day (275 persons, 2165 ms).

### 6.3. Errors analysis

The participants made 846 errors altogether, which accounts for 1.7% of all executed orders. A nonparametric, statistical test Chi-square was employed to verify the significance of differences in observed incorrect selections for various factors. The biggest proportion of mistakes (1.8%) was noted for subjects within the age range 21–25 years old, while the lowest number of errors was registered among participants, who were in their forties or older ($\chi^2 = 13.3$, $df = 5$, $p = 0.0206$). It was also observed that men made decidedly more incorrect selections (2.1%) than women (1.3%). This difference was statistically meaningful ($\chi^2 = 49.2$, $df = 1$, $p < 0.0001$). The effects of panel orientation, location, arrangement, and icon sizes did not influence the number of errors made.

# 7. Model development

## 7.1. Variables selection

The mean acquisition time (*MT*) in milliseconds computed for individual sets of buttons was taken as a dependent variable. The following independent variables were taken into consideration during model building:

*Size* − the size of graphical elements included in examined panels. Because all studied structures contained square objects, a side width in TWIPs was taken as an item size.

*Hor* − horizontality defined as a ratio of panel width to its height.

*Dis* − panel dispersion specified as a quotient of the longer side of the graphical structure to its shorter edge.

*Ver* − verticality computed by dividing panel height by its width.

*Con* − concentration expressed by a quotient of the shorter side of a given panel to its longer dimension.

*LLoc* − panel location on the left, if the examined layout was positioned on the left hand side of the screen this binary variable took one otherwise zero value.

*RLoc* − panel location on the right, if the tested structure was situated on the screen right hand side this binary variable took one otherwise zero value.

*HoAr* − horizontal arrangement, two-state variable taking one for every horizontal panel configuration.

*SqAr* − square arrangement, binary variable specifying whether a given panel has a square shape − value one, or not − value zero.

*VeAr* − vertical arrangement, variable taking one for all vertical structures and zero for all the others.

*Err* − per cent of errors made in relation to all selections for the given panel.

*Cen* − the distance between the centre of the START button and the panel centroid measured in TWIPs.

*ID* − index of difficulty calculated according to the Fitts' law formula[*] proposed by Welford (1960) with the value of movement amplitude (*A*) equal to the *Cen* variable.

Additionally, squares and cubes of *Size*, *Hor*, *Dis, Ver*, *Con, Cen*, and *ID* as well as some products of two or three of these variables were also included. We focused on those interactions that could have a significant impact in light of the analysis of variance results.

The variable selection was conducted with the aid of the following regression methods: forward and backward stepwise, forward entry, backward removal, and best subsets. Taking advantage of these procedures, the several best model variants were chosen. Additional variable selection criteria was taken into account such as, maximization of $R^2$ coefficient with as small as possible a number of dependent variables in a model, inclusion of only those variables that had as large as possible values of semi partial correlations, and avoidance of including variables that correlated strongly with each other.

## 7.2. Model formula

A regression model was constructed using the itinerant techniques of variable selection and taking into account the criteria presented above. The regression model estimated model parameters by means of the least square method the model and took the following form:

$$MT = 3361 - 4.56 \cdot Size + 0.00399 \cdot Size^2 - 19.6 \cdot Hor + 26 \cdot Dis$$

---

[*] $ID = \log_2(A / W + 0.5)$.

According to the above model, the panel mean acquisition time depends on the size of the objects included in the sets (*Size*, *Size*$^2$), the degree of panel horizontality (*Hor*), and dispersion (*Dis*). All of the model factors, including a free expression (intercept), had statistically significant influences on the dependent variable. A *t*-Student test was used to verify if the model parameters differ from zero. Results of these tests demonstrated that hypotheses about insignificance of individual parameters and the intercept value should be rejected ($p < 0.015$). The independent variables explain 90.8% of the data, while the adjusted determination coefficient amounted to $R^2_{Adj} = 88.5\%$. The employed *F*-Snedecor test proved that $R^2$ considerably differs from zero $F(4, 16) = 39.4$, $p < 0.00001$. Mean acquisition times gathered for individual graphical structures during experiments, along with values calculated from the constructed model, are illustrated in Figure 12.

Coefficients of correlations between model variates are placed in Table 10. Variables characterizing icon sizes in examined sets (*Size* and *Size*$^2$) correlated with the dependent variable the most whereas, the smallest values of correlation had the degree of panel horizontality. The highest correlations among explanatory variables occurred between *Size* and *Size*$^2$. The structure horizontality (*Hor*) and the degree of dispersion (*Dis*) did not correlate with *Size* and *Size*$^2$ variables. The correlation between *Hor* and *Dis* is caused by the way these components are defined. The *Hor* specifies the degree of panel configuration horizontality calculated as (panel width or number of columns) / (panel height or number of rows) while *Dis* is computed according to the formula (panel longer side) / (panel shorter edge). Such definitions signify that *Hor* and *Dis* take identical values for squares and horizontal configurations. In the case of vertical arrangements the horizontality value is always lower than one, whereas the dispersion rises together with the number of rows in a panel. It should also be noted that the lowest possible value of dispersion is equal to one, while any other rectangle whether it is vertical or horizontal in shape, is characterized by a bigger *Dis* value.

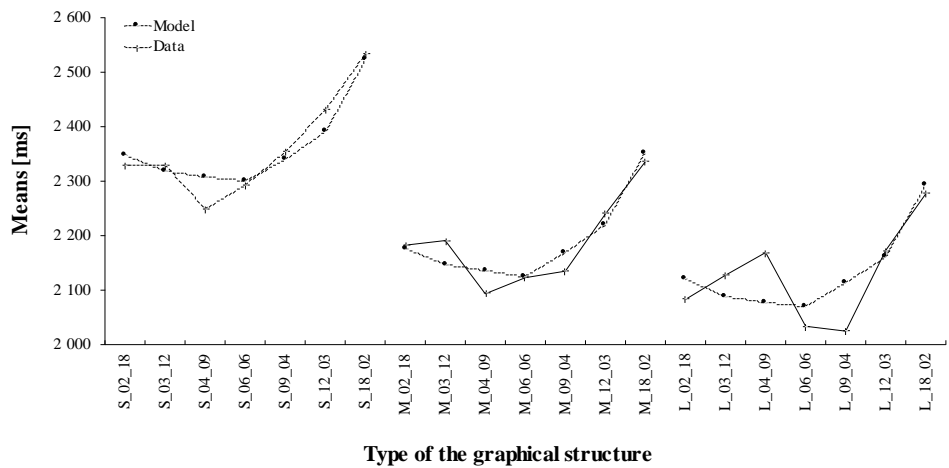

Figure 12. Mean acquisition times calculated from a model and registered during experiments (S –small, M – medium and L – large item sizes).

Table 10. Correlation matrix for the model variables.

|        | Size  | Size$^2$ | Hor   | Dis  | MT    |
|--------|-------|-------|-------|------|-------|
| **Size**   | 1     | 0.997 | 0     | 0    | −0.74 |
| **Size$^2$** | 0.997 | 1     | 0     | 0    | −0.72 |
| **Hor**    | 0     | 0     | 1     | 0.47 | −0.17 |
| **Dis**    | 0     | 0     | 0.47  | 1    | 0.4   |
| **MT**     | −0.74 | −0.72 | −0.17 | 0.4  | 1     |

In order to determine the real influence of particular predictors on the dependent variable, it is appropriate to use squared semi partial correlations (*SSC*) of individual factors with the response variable[*]. The smallest values of *SSC* had variables concerning the element's sizes ( $SSC_{Size} = 0.071$ and $SSC_{Size^2} = 0.044$ ). The situation is brought about by a very high correlation coefficient between *Size* and *Size$^2$* variables (Table 10). If the *Size$^2$* variable was removed from the model, then $R^2$ would decrease by merely 4.4% because the substantial part of the variability of the dependent variant would still be explained by the *Size*. However, when the *Size$^2$* is not included in the model, the squared semi-partial correlation for *Size* amounts to 54.7%, whereas the *SSC* for the remaining variables does not change. The greatest impact for explaining the variability of the response variable is the dimension of graphical objects included in examined panels; the second most important component is the degree of structure dispersion ( $SSC_{Dis} = 0.29$ ), and the least important is the horizontality of analyzed configurations ( $SSC_{Hor} = 0.16$ ).

### 7.3. Verification of model assumptions

The least squares method of parameter estimation gives the best possible results when appropriate assumptions are met. Initial verification of the random component normality, stationarity, and symmetry, with respect to zero, was carried out by means of graphical techniques. The normal probability plot for residuals demonstrated in Figure 13 gives grounds for stating that the random error had a normal distribution because residual values are not far away from the outlined straight line. The standardized residuals graph presented in Figure 14 does not provide a basis for rejecting the assumption about random location of residuals.

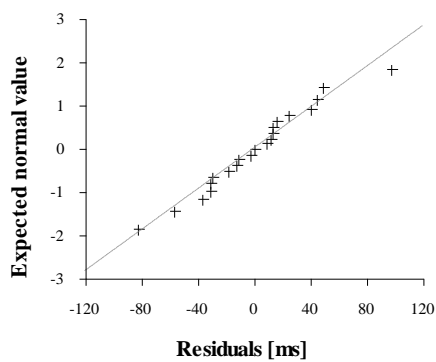

Figure 13. Normal probability plot for residuals.

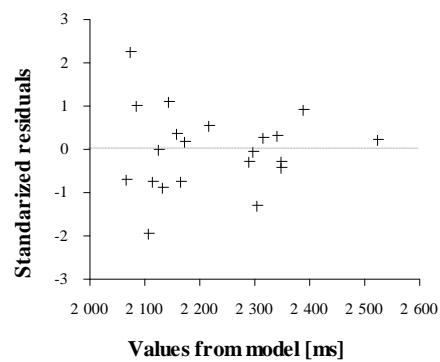

Figure 14. Plot of standardized residuals against values computed from the model.

---

[*]Squared semipartial correlation says by what value will $R^2$ be smaller if a given variable is removed from the model. Taking advantage of this feature it can be specified, which of the dependent variables has the biggest share in the model determination coefficient value.

The random error stationarity assumption was additionally verified with the aid of the Szroeter test (Dielman 2001). The value of calculated statistics amounted to $Q = 1.55$, $p = 0.120$, which did not allow us to reject the hypothesis regarding the random component constant variance at $\alpha = 0.05$. The Kolmogorov-Smirnoff test with Lilliefor's correction did not give a basis for rejecting the hypothesis about normal distribution of examined residuals ($\alpha = 0.2$). The Durbin–Watson test was employed for verification of whether the first order autocorrelation of random component occurs. Because for the studied model the computed Durbin–Watson statistics ($d = 2.04$) is considerably larger than the value of $d_U = 1.67$ ($n = 21$, $k = 3$, $\alpha = 0.05$), there was no sufficient evidence to reject the null hypothesis about the lack of random component autocorrelation. In order to examine random error symmetry, the following hypothesis was used: $\dfrac{m}{n} = \dfrac{1}{2}$ ($m$ – number of observations in plus, $n$ – number of all observations). The result of the appropriate statistics did not give a reason to discard the null hypothesis regarding symmetrical location of a random component with regards to zero ($t = 0.213$, $df = 20$, $p = 0.833$).

In order to detect multi-correlations, the variance inflation factors ($VIF$) were used. Because none of the calculated $VIF$ values ($VIF_{Size} = 1$, $VIF_{Size^2} = 1$, $VIF_{Dis} = 1.28$, and $VIF_{Hor} = 1.28$) were larger than the threshold value of 10 (Dielman 2001), the multi-correlations did not have a negative influence on the least squares estimation results. The variables $Size$ and $Size^2$ strongly correlated with each other (Table 10) due to their qualitative identity, and for that reason, during the calculation of the $VIF_{Size}$ the $Size^2$ variate was not taken into consideration; in case of $VIF_{Size^2}$, the Size variable was also not included. The mean acquisition times calculated for all analyzed panels were included in the model because during the analysis no outliers or other influential values were observed.

## 8. Discussion

The purpose of this study was to investigate the effects of some graphical interface characteristics on human-computer interaction task efficiency. In general, the presented results indicate that in the case of panels that are visible on the computer screen only during the selection process, the geometrical factors of the graphical structures play an important role in the context of operation efficiency. However, these factors have no statistically significant impact on the quantity of incorrect selections. Thirty six objects for each configuration were utilized in this study, which is substantially more than in similar experiments conducted by Grobelny and colleagues (8 and 16 icons). In other research related both to movement time and visual search in graphical interfaces, the number of items varied from six (Fleetwood and Byrne 2002, 2003; Everett and Byrne 2004, Fleetwood and Byrne in press) up to thirty (Simonin and colleagues 2005).

### 8.1. ANOVA results

#### 8.1.1. Item size

Procedures verifying a hypothesis about the graphical objects' size influence showed that the bigger the square elements, the shorter the operation time for a given panel. The sets with large objects were operated, on average, 2.7% faster than those containing the medium sized icons. The mean acquisition times for the medium size configurations were 7% higher compared to those observed for groups comprising of small objects.

Although the methods used in our study differ considerably from those used in other studies, our findings are generally in agreement with results obtained by Näsänen and colleagues (2001a). They observed for all conditions the decrease in search time with the increasing character sizes. Our results also confirm outcomes from studies on different software keyboards sizes e.g. in the work of Sears and colleagues (1993) bigger keys' sizes resulted in shorter task execution times both for novices and experienced users. However, in some research, especially when the stylus was used as a pointing device, the keyboard size had no significant effect on the task completion speed (e.g. MacKenzie and Zhang 2001, Sears and Zha 2003). Ambiguous results were obtained by Grobelny and colleagues (2005). They reported that for panels containing 8 icons, the largest items were operated 9.6% faster than the medium ones and the speed of selecting medium-sized objects was, on average, higher by 4.2% than the small ones. For structures consisting of 16 buttons, the medium panels were the best in terms of the operation speed. They were faster by 8.3% than the smallest and 4.6% than the largest icons. These results, however, are not contradictory because Grobelny and colleagues (2005) utilized different object sizes than we had. The size of small, medium and large items in their work equalled in visual angles 55', 83' and 110' respectively, while in current study the sizes were set at 41', 55', 69'.

## 8.1.2. Panel configuration

The arrangement of tested graphical structures had a significant impact on their operation efficiency. The sets in configurations of nine rows and four columns (2.12 s), as well as squares (2.13 s), were operated the quickest. The worst times were acquired for arrangements of vertical orientation, consisting of two columns and 18 rows (2.34 s) coupled with three columns and 12 rows (2.22 s).

In order to make at least rough comparisons with other studies, we also analyzed orientation effects (despite the unbalanced ANOVA). We found that mean operation times were decidedly shortest for square shape panels, while on average, far more time was needed to select elements from sets oriented vertically. Earlier studies provide inconsistent results in this respect. In the work of Deininger (1960), the task performance differences were not statistically meaningful among the best configurations including horizontal, vertical, and square arrangements. No effect of configuration was also reported by Simonin and colleagues (2005) however in this case, only one configuration had a rectangular shape – five rows and six columns. Apart from that, in all display layouts the whole screen space was used, objects were located all over the screen. In turn, Backs and colleagues (1987) reported that finding a target object and reporting an associated numerical value in menus was significantly faster in vertical than in horizontal configurations. It should be noted that in these experiments horizontal layouts could be confusing for users, as in many cases the value associated to a given menu item was located closer to the adjacent menu text. This result was not supported in the research of Scott and Findlay (1991), who showed the superiority of horizontal configuration (25 characters high and 100 wide) over the vertical one (100 high and 25 wide). Also, Shih and Goonetilleke (1998) confirmed that horizontal menus were more efficient than vertical irrespective of whether Chinese or English language was used. Similar outcomes were reported by Pearson and Schaik (2003) who demonstrated that horizontal menu locations (top and bottom) in mock web pages were searched significantly faster than vertical ones (left, right). Schaik and Ling (2001) used the same menu locations as Person and Schaik and obtained better results for menus located at the top or left of the screen. The difference could be caused by the types of link words used in both investigations. Pearson and Schaik used unrelated link words that were matched for word frequency, while Schaik and Ling utilized words that represented common concepts without matching for word frequency. Designing menus in web pages with related words could be more realistic but, in general, considering the unrelated concepts seem to be more appropriate. In experiments carried out by Grobelny and colleagues (2005), the square arrangements were operated considerably faster and the horizontal structures were better than vertical in terms of efficiency, both for the panels with 8 and 16 icons. Obtained in current study results fully confirm a relation reported by Grobelny and colleagues.

### 8.1.3. Panel location

The panel location factor was meaningful only in the case of vertically oriented structures. This situation could be caused by the fact that for medium and large elements grouped horizontally, the panels occupied more than half of the screen width. As a result, changes in their screen position modified the experimental conditions only slightly. It is possible that when a smaller number of icons were used the discrepancies between horizontal layouts would become more distinct. Of course, this hypothesis should be experimentally verified. Additionally, among the vertical panels only arrangements 09_04 and 18_02 were more quickly operated on the right hand side of screen whereas, for configurations with three columns and twelve rows the more efficient panels were positioned on the left hand side of the screen.

Previous research results regarding the effect of control items location on operation efficiency were also ambiguous. Schaik and Ling (2001) demonstrated that left or top position of navigation menu were superior to web site layouts with menus situated on the right or bottom page side in relation to correctly identified targets, while there was no difference with correct rejections. However, in subsequent studies Pearson and Schaik (2003) in a very similar experiment showed no difference between left and right hand side position of the menu, both for correctly identified targets and correct rejections. The significant discrepancies between top and bottom locations were not observed either.

### 8.1.4. Errors

The total number of errors did not exceed 1.7% on any of the trials and is generally consistent with other studies. The accuracy obtained in the work of Backs and colleagues (1987) exceeded 97%. In Scott and Findlay (1991) investigation 4.8% of visual search tasks exceeded the time limit of 180 seconds. Above 97% of correct responses for all experimental conditions were demonstrated by Shih and Goonetilleke (1998). Schaik and Ling (2001) reported that the mistakes in correctly identifying search targets equalled 1.6%, while the overall percentage of an incorrect response to a target word that was absent from the web page menu amounted to 3.11%. The error rate in correct rejections obtained by Pearson and Schaik (2003) added up to 2.77%, whereas overall incorrect responses rate to a target word present in the menu equalled to 11.99%. Task completion failures rate in experiment of Simonin and colleagues (2005) amounted to 7.5%, while errors reported in the research of Grobelny and colleagues (2005) occurred only in small and middle-size icon sets, giving less than 2% rate for all experiments. Some of the described differences can be attributed to considerably different methods employed in these individual studies.

### 8.2. Model

The constructed formal model specifies mean acquisition time for examined graphical structures as a function of the side length of objects *(Size)*, along with the degree of panel horizontality (*Hor*) and dispersion (*Dis*). The quality of the fit coefficient $R^2$ amounted to 91% (adjusted 89%). Presence of the variate, a square of the object side length ($Size^2$), suggests that in the case of different object shapes, the *MT* would probably be dependent on the item area. There was a substantial correlation between the *Hor* and *Dis* variables (0.47) resulting from their definitions. These variables appear to be almost identical, when in fact they describe quite different geometrical features of a panel. The presence of both these variates in a model is justified by the employed statistical procedures and due to the possible explanation related to oculomotor activities that are presented in the explanatory factors section. It should be noted that the employed variable selection procedures do not allow for the inclusion of factors concerning the panel location on the screen. Furthermore, the model does not contain variables connected with the number of errors made and the categorical factors related with panel orientation.

The experimental procedures in this study allowed for an element to appear more than once when testing the same configuration. This decision was dictated by making our experiments more close to the real tasks, in which a user sometimes needs to select the same control object. It was possible that memory of the location could make the next selection of the same item quicker, so to examine the impact of multiple appearence of the same execution orders for one participant in an individual layout, the obtained results were reconsidered without the recurring items. The data concerning objects that were selected more than once by a given person in the same panel were replaced by one database record with the longest acquisition time registered for the specific item. Although approximately 11% of the recorded selections were excluded, the ANOVA results were almost identical to those reported in the Results section. The results were also checked if the proposed model and its assumptions were sustained. The final conclusion was that all model parameters remained significant, and the determination coefficient amounted to $R^2 = 91.1\%$ (adjusted 88.9%) being even slightly higher than this from original model ($R^2 = 90.8\%$, adjusted 88.5%). Additionally, none of the assumptions were rejected and there were no considerable changes in semipartial correlations.

In previous HCI research most of the proposed models were connected with motor activities and usually employed various versions and modifications of Fitts' law. Cognitive factors were not considered in these models. We checked to what extent the Fitts' law would predict the mean operation time, and using the Welford (1960) formula we obtained for all experimental conditions the following equation, $MT = 1264 + 230 \times ID$, with the index of performance $IP = 5.50$. The amplitude ($A$) was calculated as the mean value of the *Cen* variable for all of the panels, while the target width ($W$) of the average item was taken. The parameters of the model were significant ($p < 0.0003$) and the fit coefficient amounted to $R^2 = 50.1\%$, (adjusted 48.3%). This result is decidedly smaller than determination coefficients usually obtained in classical Fitts' experiments, but similar to outcomes reported by Grobelny and colleagues (2005): $R^2 = 49\%$ for 8 items panels ($p = 0.036$, $IP = 1.90$) and $R^2 = 40.6\%$ in the case of 16 icons ($p = 0.065$, $IP = 9.70$). Grobelny and colleagues (2005) also proposed an overall regression equation to all of their four experiments using the factors of $ID$, $Mode^*$, and the the the number of icons ($H$). However, as far as our experimental conditions are concerned, *Mode* and $H$ are constants so the regression equation is reduced only to the $ID$ variable and takes the form of classical Fitts' equation.

In order to analyze the suitability of the Hick–Hyman law[†] (Hick 1952, Hyman 1953) in the task of searching and selecting items from rectangular structures we used overall mean time from this study, mean acquisition times for panels containing 8 and 16 items from the research of Grobelny and colleagues, and a movement time for one object calculated by means of the Fitts' formula for computer tasks[*], presented in the work of MacKenzie (1992). The obtained simple regression took the form $RT = 445 + 342 \log_2(n)$, and the fit coefficient amounted to 98.98% (adjusted 98.47%). The model was significant ($p = 0.005$) and the parameters were considerably different from zero ($p < 0.05$). Although the methods used in this research differ from those used by Grobelny and colleagues, the Hick–Hyman law was supported.

---

[*]This categorical variable took 1 if the panel was visible all the time during experiments and a value of 2 – when the icons were shown only if a specific action was taken by the participant. In our studies we focused on the latter mode.

[†]The Hick–Hyman equation relates the reaction time ($RT$) to the set of $n$ stimuli and takes the following form $RT = a + b \log_2(n)$, where $a$ and $b$ are empirically determined constants.

### 8.3. Explanatory factors

It was shown in this research that the size of the items included in the analyzed graphical structures had the greatest effect on explaining the variability of the observed mean acquisition times. Because the proposed model's statistical procedures did not allow for inclusion of the *ID* variate, it is highly probable that the movement and search times were overlapped thus the total acquisition times could be to a larger degree affected by other factors. The assumption of interfering of these two processes is supported by observations of participants' behavior made during experiments and qualitative analysis of mouse movement trajectories recorded during preliminary experiments in our ongoing investigations. In both cases it appears that users generally move the mouse pointer straight towards the tested panel relatively fast while simultanously searching for the target. The movement slows down considerably when the person is focused on finding the given item. In this phase the mouse pointer is either motionless or is being moved slowly and irregularly over the available objects finally reaching the desired target. Therefore, it can be assumed that *Size* variable, which had the biggest share in explaining the mean operation times in the proposed model, is related with the *ID* from the Fitts' law only to a small degree and affects more the search time.

According to the model, the bigger item size, the faster the panel was operated and this relation was approximated by a part of the parabola. If other variables (*Hor*, *Dis*) are constant, the extrapolation of the curve could confirm existence of optimal icon size in the panels consisting of 16 icons reported by Grobelny and colleagues (2005), all the more that our largest objects were decidedly smaller than the biggest icons used in their studies. Since the number of objects in the present study was constant, it is obvious that panel information densities defined as the quotient of information amount and the panel area are strictly correlated with item sizes. Thereby the possible inverted U-shaped function relating panel operation efficiency with the panel information density confirms conclusions drawn by Tullis (1983) after the review and analysis of studies on visual search performance and display density of alphanumeric displays.

The configuration horizontality factor effect obtained in our study results can be to some extent explained by the culturally conditioned habit of reading and scanning objects horizontally from the left to the right hand side. This effect could be additionally heightened by the use of icons representing only numbers and characters. However the results obtained by Shih and Goonetilleke (1998) suggest that results of this investigation may also be appropriate for Chinese users. The practice of using horizontally shaped menus as well as toolbars in contemporary computer systems as defaults could have a positive impact on the operation efficiency, as most of the participants were advanced computer users. The advantage of horizontal panels may result from the shape the human's field of view which is limited with its shape similar to a horizontal ellipse. Although for various wave lengths of the visible electromagnetic field the view area has a slightly different shape, the width is always larger than the height.

---

[*] *Movement time* = $12.8 + 94.7 \times ID$, where the *ID* was the average value of *ID*s calculated for all structures tetsted in the present study. The movement time amounted to 405 milliseconds.

The physical structure of the human retina and particularly its very small central area presumably underlie the faster search times for compact layouts. The foveal region is not flat as the rest of the retinal surface, but has the shape of shallow pit which is responsible for acute vision (Findlay and Gilchrist 2003). According to Polyak (1957), the pit diameter corresponds to 5 degree of visual angle and because of the fovea construction the visual acuity gradually declines as the visual angle from the centre of focus increases. At the distance of 2.5 degrees from the point of fixation it falls off approximately by half. This phenomenon was reported for the first time by Wertheim (1894) and confirmed by numerous subsequent studies. In the present study, dimesions of square configurations only slightly exceeded the 5 degree visual angle (Table 4) and hence, required probably less number of fixations to find the target than was necessary in the case of vertical or horizontal panels. Therefore, this anatomical feature of the sight organ could make searching compact configurations much easier than searching dispersed ones.

## 8.4. Limitations

This research substantially precluded novice users that would have experienced interfaces that were not known to users e.g. they use them for the first time or sporadically. The results relate also to panels (sets of options, dialogue windows etc.) which are visible on a screen only during the selection process. Neither users preferences of various types of interfaces nor other than 'point and click' with a mouse methods (e.g. command lines, drag & drop) are taken into account in this study. We have used square buttons with Latin characters and Arabic numbers, so it is possible to obtain different results using other graphical objects and button shapes. The research was confined only to 21 various rectangular panel configurations, therefore it is not possible to make conclusions based on the proposed model about different types of graphical structures, such as u-shaped panels for instance. The data from demonstrated experiments were collected in the laboratory environment on a relatively homogenous group of participants – the majority young and having substantial experience in operating computer programs.

## 8.5. Future research

The results analyzed in this study are concerned with specific graphical items presenting exclusively Latin characters and Arabic numbers. In reality, objects containing such icons appear in computer applications comparatively rarely and are usually used in software simulated keyboards. Thus further research should include other icons e.g. abstract geometrical objects or items used in contemporary interacting systems. Additional experimental research is needed to clarify obtained ambiguous results concerning the impact of panel location (left and right) on mean acquisition times. There are, of course, some other issues that should be explained in this type of interaction, for example: performance of beginners and advanced computer users, intercultural differences, and other panel locations.

It may also be verified if  these study results are also valid for other pointing devices or for touchscreens. Other areas of interest may cover sequential visual search tasks in panels of various graphical characteristics. Moreover, it is also very interesting to learn which of the panel structures are better perceived by users especially in those cases where there were insignificant differences in operation efficiency. Then guidelines for designers could be based on subjective ratings. As it was mentioned in the Experimental design section we plan to take advantage of AHP to obtain subjective opinions, and conjoint methodology to determine the conformity between objective measures (acquisition time and error rates) and vectors of participants' preferences. And finally, in order to broaden the analysis one may take advantage of eye tracking systems, fMRI[*] devices or mouse pointer trajectories registered during experiments.

## 9. Conclusions

The presented studies confirm that geometrical factors significantly affect operational efficiency for a group of control elements in the visual interactions of a human-computer interface. The effect of panel location occurred to have no impact on operation efficiency, so it in this case other factors may be consider during the design process, such as users' preferences or developer's convenience. The obtained results and conducted analysis contribute to understanding human behavior in 'point and click' method of communication, and allow for making some recommendations for interface designers. This research outcomes imply avoiding small graphical items in computer interface design, however it should be noted that taking into account other studies there could exists an optimum of item size. Additionally, increasing the object dimensions is not always the best solution, as excessive large panel buttons may, in some cases, 'steal' space meant for performing specified activities. For instance, in CAD software systems, the size of area used for drawing can be considerably decreased by a huge number of icons, even to the extent in which it would be very difficult to work efficiently. It was also shown here that graphical structures having a very high degree of concentration were operated the fastest. If square configurations cannot be used, it is advisable to apply horizontal panels that are as compact as possible. Evidently the worst, as regards to operation efficiency, were vertical arrangements, so they should not be included in designed interfaces at all.

It may be argued whether the differences detected in our studies – though statistically significant – are of any practical significance. The majority of computer applications enlist the visual search and selection of graphical objects from a group of controls, so the benefits from improving the efficiency of software operation only by a fraction on a task level, but may be substantial in a global scale. The cost of making appropriate changes in interfaces during the design phase is small and in some cases, the panels' configurations and icon sizes may be adjusted by altering existing software parameters. Similarly, small differences may become crucial in interfaces used in e.g. military applications, energy plants or for controlling surgery robots. Furthermore, our results are based on experiments carried out in a laboratory, so in the real life the differences may become larger due to user stress, lack of attention or operator fatigue, etc. Additionally, making the graphical interface operated faster may improve the perception of the given software and thus could be a decisive factor of winning with the rival products on the market.

## Acknowledgements
The authors would like to acknowledge the valuable assistance of Dr. John Layer in editing the final version of the paper.

---

[*]Functional Magnetic Resonance Imaging.

## References

BACKS, R., WALRATH, L. and HANCOCK, A., 1987, Comparison of horizontal and vertical menu formats. In *31st Annual Meeting of the Human Factors Society* (Santa Monica: Human Factors and Ergonomics Society), 715–717.

BALAKRISHNAN, R., 2004, 'Beating' Fitts' law: Virtual enhancements for pointing facilitation. *International Journal of Human-Computer Studies*, **61**, 857–874.

BLANCH, R., GUIARD, Y. and BEAUDOUIN-LAFON, M., 2004, Semantic pointing: Improving target acquisition with control-display ratio adaptation. In *CHI 2004: Human Factors in Computing Systems* (New York: ACM), 519–526.

CARD, S.K., ENGLISH, W.K. and BURR, B., 1978, Evaluation of mouse, rate-controlled isometric joystick, step keys, and text keys for text selection on CRT. *Ergonomics*, **21**, 601–613.

CARD, S.K., MORAN, T.P. and NEWELL, A., 1983, *The psychology of human-computer interaction* (Hillsdale: Lawrence Erlbaum Associates Inc.).

CHRISTOU, G. and JACOB, R.J.K., 2003, Evaluating and comparing interaction styles. In *DSV-IS 2003 10th Eurographics Workshop on Design, Specification and Verification Interactive Systems*, J.A. JORGE, N.J. NUNES, and J.F. CUNHA (Eds.) (Berlin: Springer–Verlag), 406–409.

CONINX, K., REETH, F.V. and FLERACKERS, E., 1997, 2D human-computer interaction techniques in immersive virtual environments. *Computer Networks and ISDN Systems*, **29**, 1685–1693.

DEININGER, R.L., 1960, Human factors engineering studies of the design and use of pushbutton telephone sets. *Bell System Technical Journal*, **XXXIX**, 995–1012.

DIELMAN, T.E., 2001, *Applied regression analysis for business and economics*. (Pacific Grove: Duxbury).

DIX, A., FINLAY, J., ABOWD, G.D. and BEALE, R., 2004, *Human-Computer Interaction*. (Harlow: Pearson Education).

DRURY, C.G. and HOFFMANN, E.R., 1992, A model for movement time on data-entry keyboards. *Ergonomics*, **37**, 129–147.

ENGLISH, W.K., ENGELBART, D.C. and BERMAN, M.L., 1967, Display selection techniques for text manipulation. *IEEE Transactions on Human Factors in Electronics*, **HFE-8**, 5–15.

EPPS, B., 1986, Comparison of six cursor control devices based on Fitts' law models. In *30th annual meeting of the Human Factors Society* (Santa Monica: Human Factors Society), 327–331.

EVERETT, S.P. and BYRNE, M.D., 2004, Unintended effects: varying icon spacing changes users' visual search strategy. In *CHI 2004: Human Factors in Computing Systems* (New York: ACM), 695–702.

FINDLAY, J.M. and GILCHRIST, I.D., 2003, *Active vision. The psychology of looking and seeing* (New York: Oxford University Press).

FITTS, P.M., 1954, The information capacity of the human motor system in controlling the amplitude of movement. *Journal of Experimental Psychology*, **49**, 389–391.

FITTS, P.M. and PETERSON, J.R., 1964, Information capacity of discrete motor responses. *Journal of Experimental Psychology*, **67**, 103–112.

FLEETWOOD, M.D. and BYRNE, M.D., 2002, Modeling icon search in ACT-R/PM. *Cognitive Systems Research*, **3**, 25–31.

FLEETWOOD, M.D. and BYRNE, M.D., 2003, Modeling the visual search of displays: A revised ACT-R/PM model of icon search based on eye-tracking and experimental data. In *5th*

*International Conference on Cognitive Modeling*, F. DETJE, D. DÖRNER and H. SCHAUB (Eds.) (Bamberg: Universitas–Verlag Bamberg), 87–92.

FLEETWOOD, M.D. and BYRNE, M.D., in press, Modeling the visual search of displays: A revised ACT-R/PM model of icon search based on eye tracking data. *Human Computer Interaction*, Available online at: http://chil.rice.edu/research/pubs.html (accessed 18 November 2005).

GOLDBERG, J.H. and KOTVAL, X.P., 1999, Computer interface evaluation using eye movements: methods and constructs. *International Journal of Industrial Ergonomics*, **24**, 631–645.

GREENSTEIN, J.S. and ARNAUT, L.Y., 1988, Input devices. In *Handbook of Human-Computer Interaction*, M. HELANDER (Ed.) (Amsterdam: Elsevier Science Publishers B.V), 485-519.

GROBELNY, J., KARWOWSKI, W. and DRURY, C., 2005, Usability of Graphical icons in the design of human-computer interfaces. *International Journal of Human-Computer Interaction*, **18**, 167–182.

GUIARD, Y. and BEAUDOUIN-LAFON, M., 2004, Target acquisition in multiscale electronic worlds. *International Journal of Human-Computer Studies*, **61**, 875–905.

HARTSON, H.R., 1998, Human-computer interaction: Interdisciplinary roots and trends. *The Journal of Systems and Software*, **43**, 103–118.

HICK, W.E., 1952, On the rate of gain of information. *Quarterly Journal of Experimental Psychology*, **4**, 11–36.

HOFFMANN, E.R. and LIM, J.T.A., 1997, Concurrent manual-decision tasks. Ergonomics, **40**, 293–318.

HYMAN, R., 1953, Stimulus information as a determinant of reaction time. *Journal of Experimental Psychology*, **45**, 188–196.

KROEMER, K.H.E., 2001, Keyboards and keying. An annotated bibliography of the literature from 1878 to 1999. *Universal Access in the Information Society*, **1**, 99–160.

LEE, P.U.J. and ZHAI, S., 2004, Top-down learning strategies: Can they facilitate stylus keyboard learning? *International Journal of Human-Computer Studies*, **60**, 585–598.

LINDBERG, T. and NÄSÄNEN, R., 2003, The effect of icon spacing and size on the speed of icon processing in the human visual system. *Displays*, **24**, 111–120.

LUCE, D.R., and TUKEY, J.W., 1964, Simultaneous conjoint measurement: a new type of fundamental measurement. *Journal of Mathematical Psychology*, **1**, 1–27.

MACKENZIE, I.S., 1991, Fitts' law as a performance model in human-computer interaction. PhD thesis, University of Toronto.

MACKENZIE, I.S., 1992, Fitts' law as a research and design tool in human-computer interaction. *Human-Computer Interaction*, **7**, 91–139.

MACKENZIE, I.S., 1995a, Movement time prediction in human-computer interfaces. In *Readings in human-computer interaction*, R.M. BAECKER, W.A.S. BUXTON, J. GRUDIN and S. GREENBERG (Eds.) (Los Altos: Kaufmann), 483–493.

MACKENZIE, I.S., 1995b, Input devices and interaction techniques for advanced computing. In *Virtual environments and advanced interface design*, W. BARFIELD, T.A. FURNESS (Eds.) (Oxford: Oxford University Press), 437–470.

MACKENZIE, I.S. and ZHANG, S.X., 1999, The design and evaluation of a high-performance soft keyboard. In *CHI 1999: Human Factors in Computing Systems* (New York: ACM), 25–31.

MACKENZIE, I.S. and ZHANG, S.X., 2001, An empirical investigation of the novice experience with soft keyboards. *Behaviour and Information Technology*, **20**, 411–418.

MACKENZIE, I.S., ZHANG, S.X. and SOUKOREFF, R.W., 1999, Text entry using soft keyboards. *Behaviour and Information Technology*, **18**, 235–244.

MACKENZIE, I.S., 1992, Fitts' Law as a research and design tool in human-computer interaction. *Human-Computer Interaction*, **7**, 91–139.

MARTIN, G.L., 1988, Configuring a numeric keypad for a touch screen. *Ergonomics*, **31**, 945–953.

MICHALSKI, R., 2005, Komputerowe wspomaganie badań jakości ergonomicznej oprogramowania. PhD thesis, Wroclaw University of Technology.

MURATA, A. and FURUKAWA, N., 2005, Relationships among display features, eye movement characteristics, and reaction time in visual search. *Human Factors*, **47**, 598–612.

MURATA, A. and IWASE, H., 2001, Extending Fitts' law to a three-dimensional pointing task. *Human Movement Science*, **20**, 791–805.

MURATA, A., TAKAHASHI, Y. and FURUKAWA, N., 2002, Evaluation of display design based on human's eye movement. In *IEEE International Conference on Systems, Man And Cybernetics* (Yasmine Hammamet: IEEE), 587–592.

NÄSÄNEN, R. and OJANPÄÄ, H., 2003, Effect of image contrast and sharpness on visual search for computer icons. *Displays*, **24**, 137–144.

NÄSÄNEN, R., KARLSSON, J. and OJANPÄÄ, H., 2001a, Display quality and the speed of visual letter search. *Displays*, **22**, 107–113.

NÄSÄNEN, R., OJANPÄÄ, H. and KOJO, I., 2001b, Effect of stimulus contrast on performance and eye movements in visual search. *Vision Research*, **41**, 1817–1824.

NELDER, J.A. and WEDDERBURN, R.W.M., 1972, Generalized linear models. *Journal of the Royal Statistical Society A*, **135**, 370–384.

NIEMELÄ, M. and SAARINEN, J., 2000, Visual Search for grouped versus ungrouped icons in a computer interface. *Human Factors*, **42**, 630–635.

PEARSON, R. and SCHAIK, P., 2003, The effect of spatial layout of and link color in web pages on performance in a visual search task and an interactive search task. *International Journal of Human-Computer Studies*, **59**, 327–353.

PLAMONDON, R. and ALIMI, A.M., 1997, Speed/accuracy trade-offs in target-directed movements. *Behavioural And Brain Sciences*, **20**, 279–349.

PREECE, J., ROGERS, Y. and SHARP, H., 2002, *Interaction design: Beyond human-computer interaction* (New York: John Wiley & Sons).

POLYAK, S.L., 1957, *The vertebrate visual system* (Chicago: University of Chicago Press).

SAATY, T.L., 1980, *The analytic hierarchy process* (New York: McGraw-Hill).

SALLNÄS, E.L. and ZHAI, S., 2003, Collaboration meets Fitts' law: Passing virtual objects with and without haptic force feedback. In *INTERACT 2003: IFIP TC13 International Conference on Human-Computer Interaction* (Zurich: IOS Press), 97–104.

SCHAIK, P. and LING, J., 2001, The effects of frame layout and differential background contrast on visual search performance in web pages. *Interacting with Computers*, **13**, 513–525.

SCOTT, D. and FINDLAY, J.M., 1991, Visual search and VDUs. In *Visual search II*, D., BROGAN (Ed.) (London: Taylor and Francis).

SEARS, A. and ZHA, Y., 2003, Data entry for mobile devices using soft keyboards: Understanding the effects of keyboard size and user task. *International Journal of Human-Computer Interaction*, **16**, 163–184.

SEARS, A., JACKO, J.A., CHU, J. and MORO, F., 2001, The role of visual search in the design of effective soft keyboards. *Behaviour and Information Technology*, **20**, 159–166.

SEARS, A., REVIS, D., SWATSKI, J., CRITTENDEN, R. and SHNEIDERMAN, B., 1993, Investigating touchscreen typing: The effect of keyboard size on typing speed. *Behaviour and Information Technology*, **12**, 17–22.

SHIH, H.M. and GOONETILLEKE, R.S., 1998, Effectiveness of menu orientation in Chinese. *Human Factors*, **40**, 569–576.

SHNEIDERMAN, B., 1982, The future of interactive systems and the emergence of direct manipulation. *Behaviour and Information Technology*, **1**, 237–256.

SHNEIDERMAN, B., 1983, Direct manipulation. A step beyond programming languages. *IEE Computer*, **16**, 57–69.

SIMONIN, J., KIEFFER, S. and CARBONELL, N., 2005, Effects of display layout on gaze activity during visual search. In *INTERACT 2005: IFIP TC13 International Conference on Human-Computer Interaction* (Berlin/Heidelberg: Springer-Verlag), 1054–1058.

SOUKOREFF, R.W. and MACKENZIE, I.S., 2004, Towards a standard for pointing device evaluation, perspectives on 27 years of Fitts' law research in HCI. *International Journal of Human-Computer Studies*, **61**, 751–789.

TORRES, R.J., 2002, *User interface design and development* (Upper Saddle River: Prentice Hall PTR).

TULLIS, T.S., 1983, The formatting of alphanumeric displays: A review and analysis. *Human Factors*, **25**, 657–682.

WELFORD, A.T., 1960, The measurement of sensory-motor performance: Survey and reappraisal of twelve years' progress. *Ergonomics*, **3**, 189–230.

WERTHEIM, T., 1894, Über die indirekte Sehschärfe. *Zeitschrift für Psychologie und Physiologie der Sinnesorgane*, **7**, 172–187.

WHITTAKER, S., TERVEEN, L. and NARDI, B.A., 2000, Let's stop pushing the envelope and start addressing it: A reference task agenda for HCI. *Human-Computer Interaction*, **15**, 75–106.

ZHAI, S., 2004, Characterizing computer input with Fitts' law parameters – the information and non-information aspects of pointing. *International Journal of Human-Computer Studies*, **61**, 791–809.

ZHAI, S., HUNTER, M. and SMITH, B.A., 2002, Performance optimization of virtual keyboards. *Human-Computer Interaction*, **17**, 229–269.

ZHAI, S., KONG, J. and REN, X., 2004, Speed-accuracy tradeoff in Fitts' law tasks – on the equivalency of actual and nominal pointing precision. *International Journal of Human-Computer Studies*, **61**, 823–856.